\documentclass[11pt]{article}
\usepackage{graphicx}
\begin{document}
%\pagenumbering{arabic}
\bibliographystyle{unsrt}
\def\ra{\rangle}
\def\la{\langle}
\def\aao{\hat{a}}
\def\aaot{\hat{a}^2}
\def\aco{\hat{a}^\dagger}
\def\acot{\hat{a}^{\dagger 2}}
\def\ano{\aco\aao}
\def\bao{\hat{b}}
\def\baot{\hat{b}^2}
\def\bco{\hat{b}^\dagger}
\def\bcot{\hat{b}^{\dagger 2}}
\def\bno{\bco\bao}
\def\beqn{\begin{equation}}
\def\eeqn{\end{equation}}
\def\bear{\begin{eqnarray}}
\def\eear{\end{eqnarray}}
\def\cdott{\cdot\cdot\cdot}
\def\bcen{\begin{center}}
\def\ecen{\end{center}}
\title{Entanglement of bosonic modes of nonplanar molecules}
%\title{Entanglement of bosonic modes}
\author{S. Sivakumar\\Materials Science Division\\ 
Indira Gandhi Centre for Atomic Research\\ Kalpakkam 603 102 INDIA\\
Email: siva@igcar.gov.in\\
Phone: 91-044-27480500-(Extension)22503}
\maketitle
\begin{abstract}
Entanglement of bosonic modes of material oscillators is studied in the context 
of two bilinearly coupled, nonlinear oscillators.   These oscillators are 
realizable in the vibrational-cum-bending motions of C-H bonds in 
dihalomethanes.  The bilinear coupling gives rise to invariant subspaces in the 
Hilbert space of the two oscillators.   The 
dynamics of entanglement and quadrature fluctuations is studied.    
Two classes of initial conditions, 
corresponding to the case where the total energy is concentrated in one of the 
modes and the other wherein the modes share the total energy, in the 
invariant subspaces are considered. 
 The inadequacy of the known entanglement detection criteria is 
established and an inseparability criterion that is applicable to the states in 
the invariant subspaces is provided.  Possibility of generating maximally 
entangled states is indicated.
\end{abstract}
PACS: 03.67.Bg, 03.65.Ca\\
Keywords: Entanglement, Kerr couplers, local modes
\newpage
\section{Introduction}\label{secI}
   System with more than one degree of freedom has the potential to get 
entangled, a feature that is essentially quantal. No classical process has 
all the features exhibited by the correlations in the entangled quantum states. 
 Such non-classical correlations are essential for quantum teleportation, 
quantum computation, etc.  Hence it is important to identify physical systems 
wherein quantum entanglement is easily generated, controlled and measured.  A 
host of proposals and demonstrations based on NMR, ion traps, SQUID, photon 
polarization, {\em etc}  are known to generate entangled states.  These systems 
can be 
used as  gadgets to perform quantum computation \cite{nielsen, devices}. 
Recently, the entangling 
capabilities of molecules and the possibility of engineering  their evolution 
to make quantum gates have been investigated\cite{tesch, babikov}.   Molecules 
have vibrational, rotational and  electronic degrees of freedom.  Often these 
degrees of freedom are coupled.  It is precisely the coupling among the various 
degrees of freedom that is exploited to generate suitable molecular states. 
Study of such systems is important as the vibrational modes perform better as 
controllable qubits\cite{babikov}.    In addition to the aforementioned degrees 
of freedom, a nonlinear molecule like the dihalomethane  has bending 
motion.  Dihalomethanes are obtained by replacing two of the hydrogen atoms in 
$CH_4$ with halogen atoms;  $CH_2Cl_2 $ being an  example.  These molecules 
have nonplanar geometrical arrangement of atoms.   A classical model to 
understand the vibrational spectra of these molecules takes the two C-H bonds 
to be two coupled quartic oscillators.   The potential energy of the quartic 
oscillator is 
\beqn
V(x) = {1\over 2}M\omega^2x^2-\alpha x^4,
\eeqn
where $x$ measures the deviation of C-H bond length from its equilibrium value.  
Here $M$ is the "effective mass" of the C-H oscillator.    
The coefficient ($\alpha$) of the quartic term is positive. The negative 
sign for the quartic term is to account for the  
dissociation of sufficiently excited molecule.  Stretching of a C-H bond 
affects the charge distribution in the molecule.  This, in turn, alters the 
potential of the other C-H oscillator, leading to an interaction between the 
two C-H bonds.  Thus, the coupling between the bonds  is the net effect of  
mechanical and electromagnetic interactions.   A first order perturbation 
theory is performed to account for the 
interaction between the bonds.  The small parameter $\epsilon$ required for the 
perturbation theory is $\sqrt{m/M}$, where $m$ is the electron mass.  
The mass of the hydrogen 
enters the analysis as the stretching of C-H bond stretching involves 
movement of hydrogen. A detailed discussion of these aspects and a  
derivation of Hamiltonian operator are given in \cite{scott}.  
 The quantum Hamiltonian,
 obtained after making rotating wave approximation is,  
\bear\label{hamiltonian}
\hat{H}&=&\left[(\omega-{\gamma\over2})(\ano+\bno)-{\gamma\over 2}
(\ano\ano+
\bno\bno)-\epsilon(\aco\bao+\aao\bco)\right],\\
&=&\hat{H}_0+\hat{H}_n+\hat{H}_c .
\eear
This Hamiltonian describes the 
low energy vibrational states of the molecule correctly.  
The convention of setting $\hbar=1$ in the expressions is followed in this 
work. 
In calculations the actual value of the Plank's constant is used. 
The Hamiltonian $\hat{H}$ describes a bipartite system 
of two coupled, bosonic oscillators.  
The operators $\aco$ and $\aao$ are the creation and annihilation operators for 
the vibrational mode corresponding to one of the C-H oscillators.  These 
operators satisfy the bosonic commutation relation $\left[\aao,\aco\right]=I$.  
The corresponding operators for the  other  C-H oscillator are $\bco$ and 
$\bao$. 
 The bending motion is modelled by the term nonlinear in $\ano$ and 
$\bno$. The term with coefficient $\omega-\gamma/2$ is $H_0$, the free 
Hamiltonian.  The nonlinear part is denoted by $\hat{H}_n$ and its coefficient 
 is defined by $\gamma=3\alpha\hbar^2/4M^2\omega^2$. The coupling term is  
$\hat{H}_i$ which has 
 $\epsilon=\sqrt{m/M}$ as the coefficient. In the context of nonlinear 
optics this type of bilinear  coupling has models nonlinear couplers for   
electromagnetic 
fields\cite{chefles, bernstein}.   The Hamitlonian has been  used in the study
  of local modes (which are the states of molecules where energy is   
concentrated in one of the C-H bonds in the  molecule) and quantum lattice 
solitons\cite{scott}.  In the present work, the Hamiltonian $\hat{H}$ is 
analyzed to study entanglement in the coupled bosonic system.  Analysis of 
bosonic systems in the context of quantum information theory is becoming 
important\cite{bernsteinRMP}, especially after the demonstration of  
teleportation of coherent states of electromagnetic fields\cite{furusawa}.  
Theoretical studies have indicated the possibility of generating gates with 
bosonic modes and the robustness of these gates against thermal dissipation
\cite{zanardi}. \\

	The Hamiltonian in Eq.\ref{hamiltonian} is similar to that used in the 
analysis of propagation of two-mode electromagnetic fields in a medium with 
$\chi^2$ nonlinearity\cite{perina, agarwalpuri}.  The nonlinear medium  couples 
the two modes of the electromagnetic field propagating in it 
\cite{jensen}.  
An important difference between the Hamiltonian for a nonlinear coupler and 
$\hat{H}$ is that the coefficient of the  nonlinear term is positive in the 
former case while it is negative in the later.
The  Hamiltonian has been investigated in the study of systems for 
generating maximally entangled states \cite{miranowicz_jpb, leonski_jopb}, 
entanglement dynamics \cite{sanz_jpa}, control of switching modes 
\cite{ariunbold}, generation of bright entangled continuous variable states 
\cite{olsen}, wavepacket dynamics \cite{sudeesh}, nonlinear quantum scissors to
 generate finite dimensional states in systems with infinite dimensional Hilbert
 space  \cite{kowalewska}and many more interesting physical applications. 

	In the case of dihalomethanes, the Hamilotnian describes mechanical 
oscillators, that is, oscillating masses.  There are no 
external agents, like the Kerr medium in the case of nonlinear couplers,  
responsible for the coupling between the two oscillators.  The coupling is 
internal and it is not possible to switch off the coupling.  The presence of 
coupling alters the dynamics of the  
two modes corresponding to the two C-H oscillators significantly.   In the 
present work, the 
physical properties such as quadrature fluctuations and  entanglement  are 
studied.  The studies pertain to the situation when  dissipation and external 
fields are absent. Effects of the nonlinearity and the coupling  on the 
dynamics are explored in this work.  The 
results presented here correspond to typical values of $\gamma,\omega$ and 
$\epsilon$ for a class of dihalomethanes.  The relevant values are tabulated in 
Table. \ref{table:parameter}\cite{scott}.
\begin{table}[ht]\label{parameters}
\caption{Values of $\gamma,\epsilon$ and $\omega$ expressed in cm$^{-1}$ units}
	\begin{center}
		\begin{tabular}{|l||c|c|l|}
\hline
Molecule  & $\gamma$&  $\epsilon$ & $\omega$\\
\hline
$CCl_2H_2$& 127.44 & 29.54 &3020.1\\
\hline
$CBr_2H_2$& 125.45 & 32.80 &3026.8\\
\hline
$CI_2H_2$& 124.25 & 33.69 &3068.7\\
\hline
		\end{tabular}
	\end{center}
\label{table:parameter}
\end{table}
The values of the parameters are nearly equal for the three species. For the 
purpose of presenting the results, the representative values used for the 
parameters $\gamma,\epsilon$ and $\omega$ are 125, 30 and 3050 respectively, 
all expressed in units of cm$^{-1}$.   
The organization of the paper is as 
follows.   In Section \ref{secII} the special features of the Hamiltonian are 
discussed.  In Section \ref{secIV}, dynamics of entanglement is studied.  
Known entanglement detection criteria are analyzed in the present context. The 
inadequacy of these criteria is established and a new criterion suitable 
for states of the present system is given.   Results on 
the fluctuations of single-mode and two-mode quadratures are presented in 
Section \ref{secV}. In Section \ref{secVI}, it is shown that maximally 
entangled states are generated at specific instants during evolution from a 
separable state. Results are summarized in Section \ref{results}. 
%%%%%%%%%%%%%%%%%%%%%%%%%%%%%%%%%%%%%%%%%%%%%%%%%%%%%%%%%%%%%%%%%%%%%%%%%%%%%%
%%%%%%%%%%%%%%%%%%%%%%%%%%%%%%%%%%%%%%%%%%%%%%%%%%%%%%%%%%%%%%%%%%%%%%%%%%%%%%
\section{Special features of $\hat{H}$}\label{secII}

The presence of the bilinear coupling term $\hat{H}_c$ endows the Hamiltonian 
$\hat{H}$  with interesting features. A suitable basis to expand an arbitrary 
state of the oscillators is the set 
$\{\vert n,m\ra\}, n,m=0,1,2,\cdot\cdot\cdot$. The quantum numbers $n$ 
and $m$ label  the states of the oscillators corresponding to the two modes.    
In the absence of the coupling,  for $n+m=N$, then there are $N+1$ 
eigenstates of the form 
$\vert N-m,m\ra$, with   $m$ ranging from 0 to $N$. The corresponding 
eigenvalues are $(\omega-{\gamma\over 2})N-{\gamma\over 2}\left[(N-m)^2+m^2
\right]$. For odd $N$, there are $(N+1)/2$ distinct eigenvalues and the states
 are doubly degenerate.  When $N$ is even, there are $(N+2)/2$ distinct 
eigenvalues; there is one nondegenerate state $\vert N/2,N/2\ra$ and the rest 
are doubly degenerate.  When $\epsilon\ne 0$, the degeneracy is lost and the 
states $\vert n,m\ra$, except when $n=m=0$, cease to be eigenstates of 
$\hat{H}$.   
The free Hamiltonian 
$\hat{H}_0$, which is the total number operator, commutes with the total 
Hamiltonian $\hat{H}$ and  the coupling term $\hat{H}_i$. Consequently, the 
Hilbert space of the coupled system splits into disjoint, irreducible, invariant
 subspaces.    Each invariant  subspace is characterized by  total the quantum 
number $N= n+m$ and  the symbol $S_N$ is used to represent the corresponding  
invariant subspace.  The subspace $S_N$ is the span of the vectors 
$\{\vert N-m,m\ra\}_{m=0}^N$ and its dimension is $N+1$.  For instance,  
when $n+m=1$,  there are two 
possible states, namely, $\vert 0,1\ra$ and $\vert 1,0\ra$.  The span of these 
two canonical basis states forms the relevant invariant subspace $S_1$.   If 
coupling is absent the irreducible, invariant subspaces are of dimension one. 
 The nonlinear and the coupling terms  
in the Hamiltonian do not commute.  Consequently, the factorization of the 
time-evolution operator is difficult.   Nevertheless, including
 a nonlinear coupling term, namely, $\ano\bno$, allows to write the Hamiltonian 
as a sum of three mutually commuting terms\cite{sanz_jpa}. In the present work  
the discussions are limited to the bilinear coupling model.\\ 

  Another remarkable feature of the invariant spaces is that {\it the states of 
the form $\vert n,m\ra$ are the only product states in the respective    
invariant subspaces.  Every other state in a given invariant subspaces is 
entangled.}  The most general state in the invariant subspace $S_N$ is of the 
form $\sum_{n=0}^N c_n\vert n, N-n\ra$.  If this is to be a product state, then 
it should be expressible as 
$\sum_{n=0}^N c_n\vert n, N-n\ra=\sum_{k=0}^N g_k\vert k\ra_a\times
\sum_{m=0}^N f_m\vert m\ra_b$, the suffixes in the states indicating the
 corresponding modes.  Consequently, the coefficients satisfy 
$c_n=g_nf_{N-n}$.  Let the state of the bipartite system is such that at least  
two coefficients, say, $c_p$ and $c_q$ are nonzero.  The relation among the 
 coefficients implies  that $g_p, g_q, f_{N-p}$ and $f_{N-q}$ are nonzero. 
 As a result, the product state is  
$g_pf_{N-p}\vert p,N-p\ra+g_qf_{N-q}\vert q,N-q\ra+ 
g_pf_{N-q}\vert p,N-q\ra+g_qf_{N-p}\vert q,N-p\ra$.  However, the states 
$\vert p,N-q\ra$ and $\vert q,N-p\ra$ do not belong to $S_N$, their total 
quantum numbers are $N-q+p$ and $N-p+q$ respectively and not $N$.   
Hence, it is not possible to express the state with two 
nonzero $c_n$ as a product of states in $S_N$. This argument can be extended 
to states 
wherein more number of $c_n$ are nonzero to show that such states are 
entangled.   If there is only one non-vanishing $c_n$, then the 
$g_n$ and $f_{N-n}$ are the only nonzero coefficients.  The corresponding 
state $\vert n,N-n\ra$  is a product state.

 Explicit construction of the eigenfunctions in terms of the number states 
and eigenvalues of the Hamiltonian is possible by the method of number states
\cite{scott}.  However, the 
resulting expressions for the eigenstates are complicated.  
On treating the coupling as a perturbation, simple expressions for   
approximate eigenstates and eigenvalues are obtained by first order 
perturbation theory.  Since $\la N-m,m\vert\aco\bao+\aao\bco\vert N-m,m\ra$ is 
zero, the eigenvalues do not change to first order in perturbation.   Apart 
from an overall nonramlization factor, the perturbed eigenstates, denoted with 
a suffix $p$, are
\beqn\label{pertstate}
\vert N-m,m\ra_p\propto\vert N-m,m\ra+f_1\vert N-m+1,m-1\ra + f_2\vert N-m-1,
m+1\ra
\eeqn
where 
\begin{eqnarray}
f_1&=&{\epsilon\over \gamma}{\sqrt{(N-m+1)m}\over 1+N-2m},\\
f_2&=&{\epsilon\over \gamma}{\sqrt{(N-m)(m+1)}\over 1-N+2m}.
\end{eqnarray}
If 
$\vert N-2m\vert=1$, either $f_1$ or $f_2$ becomes infinity and the 
perturbation theory is not applicable. 
Though the eigenstates are doubly degenerate, the results of nondegenerate 
perturbation theory are applicable if $\vert N-2m\vert\ne1$.  This holds for 
all for states except those states of the form $\vert m,m+1\ra$ and 
$\vert m+1,m\ra$.  The transition matrix element $\la m+1,m\vert\aao\bco+
\aco\bao\vert m,m+1\ra$ is nonzero and hence the nondegenerate perturbation 
expansion is invalid.   The matrix element of $\aao\bco+\aco\bao$ among the 
other states vanish and the Eqn.\ref{pertstate} holds for such states.    
Further it is required that  $\gamma>>\epsilon$ for the perturbation expansion 
to be
valid and this condition holds in the case of dihalomethanes.  
Two important special cases,  $\vert N,0\ra$ and its degenerate 
counterpart $\vert 0,N\ra$, are analyzed.   Physically, these states 
correspond to the situation in which the total energy is concentrated in one 
of the modes and are referred as "local modes".   From the expression for 
perturbed states in Eqn. \ref{pertstate}, 
\begin{eqnarray}
\vert N,0\ra_p&\propto&\vert N,0\ra+{\epsilon\over\gamma}{\sqrt{N}
\over N+1}\vert N-1,1\ra,\\
\vert 0,N\ra_p&\propto&\vert 0,N\ra+{\epsilon\over\gamma}{\sqrt{N}\over N-1}
\vert 1,N-1\ra ,
\end{eqnarray}
where $N > 1$ to ensure that the perturbation results are valid.  
For large $N$, the second term of the perturbed state is negligible. 
Consequently, the unperturbed states 
$\vert N,0\ra$ and $\vert 0,N\ra$ qualify as approximate eigenstates of the 
perturbed Hamiltonian $\hat{H}$.  The validity of the result is ascertained by 
studying the overlap  between  the initial state and the corresponding 
time-evolved state. Fidelity is defined as the absolute value of overlap of the 
initial state with the time-evolved state. This definition means that 
the fidelity is given by the 
absolute value of the expectation value of the evolution operator 
$\exp(-it\hat{H})$ evaluated in the initial state. If a state is an 
eigenstate of 
$\hat{H}$ then its fidelity is unity at all times.  The fidelity  for the 
states $\vert 2,0\ra$ and $\vert 1,1\ra$ are given in given in Fig. 1a.    
The results are based on nonperturbative evolution of the initial state so 
that the applicability of the perturbation theory is justified.  The 
explicit construction of the Hamiltonian matrix required for the numerical 
computation is specified in the next section.  
The fidelity of the 
state $\vert 2,0\ra$ becomes very small, an indication that  the state is not 
an eigenstate.  In the same figure the fidelity of the state $\vert 1,1\ra$ is 
given.  Though the fidelity does not become zero, its deviation from unity is 
significant as the state is not an eigenstate.  In Fig. 1b fidelity of the 
states $\vert 0,4\ra$ and $\vert 2,2\ra$ are given.  It is seen that the 
fidelity of the former remains close to unity for all times.  Thus the state 
$\vert 4,0\ra$ is an approximate eigenstate.  This is expected based on the 
perturbation theory result that for large $N$ the local modes are approximate 
eigenstates. The fidelity of the state $\vert 2,2\ra$ is appreciably different 
from unity since the state is not a good approximation to an eigenstate. 
Though the state $\vert 1,3\ra$ has significant amount of energy concentrated 
in one of the 
modes, the total quantum number (equal to 4 in this case) is not sufficient 
to make it an approximate eigenstate.

The role of nonlinear term 
in the Hamiltonian deserves to be stressed.  
 For the local modes to be 
approximate eigenstates the  nonlinearity parameter $\gamma$ should be  large 
in comparison to the coupling strength $\epsilon$.   In addition, the 
eigenstates of $\hat{H}$ are all entangled, except the ground state, 
 if nonlinearity is absent or weak. 
Nonlinearity makes it possible for the system to have approximate eigenstates 
 which are separable, despite the  presence of coupling.  If $\gamma$ vanishes, 
first order perturbation theory fails and the states of the form $\vert n,m\ra$ 
do not approximate the eigenstates.  

\begin{figure}[htp]
\centering
\includegraphics[width=0.9\textwidth, height=8cm]{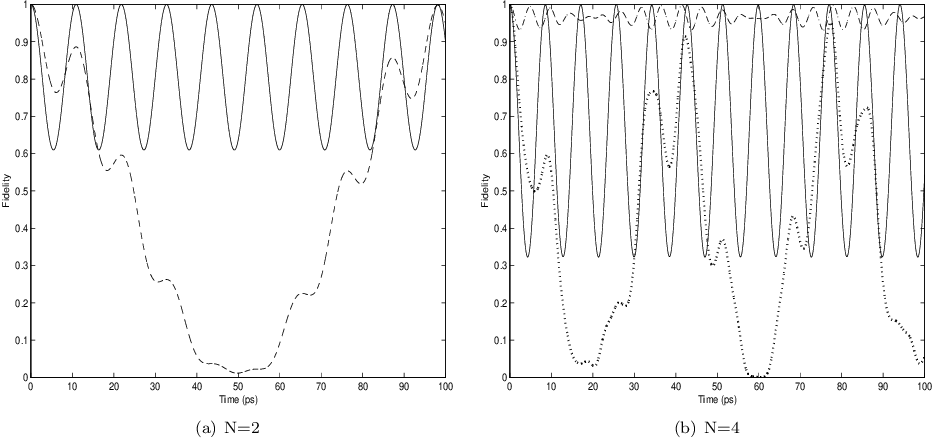}
\caption{Fidelity $\la\exp(-it\hat{H})\ra$ of states corresponding to total 
quantum numbers $N=2$ and $N=4$.  The expectation value is evaluated in the 
states for which fidelity is calculated.  Fidelity of 
$\vert 0,2\ra$ (dashed) and $\vert 1,1\ra$ (continuous) are shown in (a). 
Fidelity of $\vert 0,4\ra$ (dashed), $\vert 2,2\ra$ (continuous) and 
$\vert 1,3\ra$ (dotted) are shown in (b).  Values used for the parameters are 
$\gamma=125~\hbox{cm}^{-1}$, $\epsilon = 30~\hbox{cm}^{-1}$ and 
$\omega= 3050~\hbox{cm}^{-1}$. }
\label{fig:Fig1}
\end{figure}
%%%%%%%%%%%%%%%%%%%%%%%%%%%%%%%%%%%%%%%%%%%%%%%%%%%%%%%%%%%%%%%%%%%%%%%%%%%%%%
%%%%%%%%%%%%%%%%%%%%%%%%%%%%%%%%%%%%%%%%%%%%%%%%%%%%%%%%%%%%%%%%%%%%%%%%%%%%%%
%\section{Evolution of the system}\label{secIII}
\section{Dynamics of Entanglement}\label{secIV}

 In the absence of coupling, the Hamiltonian is a function of the two number 
operators, namely,  $\ano$ and $\bno$.  The states $\vert N-m,m\ra$ are 
eigenstates of the 
coupling-free Hamiltonian.   From the expressions for the perturbed eigenstates  
given in  Eq. \ref{pertstate}, it is clear that if the nonlinearity dominates 
over the coupling, $\gamma>>\epsilon$, the separable states $\vert N-m,m\ra$ 
become the approximate eigenstates of the total Hamiltonian.  The other extreme
 case corresponds to the absence of the  nonlinear term  while the coupling is 
retained.  Perturbation theory results are not valid in this case. However, in 
this limit the Hamiltonian $\hat{H}$ is the sum  two operators, namely, 
$\ano+\bno$ and $\aco\bao+\aao\bco$, which commute. By the 
Baker-Campbell-Hausdorff formula\cite{mandelwolf}, the unitary time evolution 
operator $\exp(-it\hat{H})$ factorises to  
\beqn
\hat{U}=\exp(-it\hat{H})=\exp(i\epsilon t(\aco\bao+\aao\bco))\exp(-i\omega t
(\ano+\bno)).
\eeqn
The time-evolved states are obtained by the action of $\hat{U}$ on the initial 
states.

 For dihalomethanes, the parameters $\gamma$ and $\epsilon$  satisfy 
$\gamma\approx 
4\epsilon$ (see Table \ref{table:parameter}). Hence, the advantages of the 
limiting 
cases, namely, $\gamma=0$ or $\epsilon = 0$, in solving for the dynamics are 
not available.  However, the availability of 
 invariant subspaces simplifies the study of dynamics if the initial state 
is in one of the invariant subspaces. If the initial state is $\sum_{r=0}^N 
c_r\vert N-r,r\ra$, the evolved state has no overlap with states in other 
invariant spaces. Hence,  the dynamics of such states is dictated by a 
truncated Hamiltonian of dimension $N+1$, where $N$ is the total quantum 
number characterizing the invariant space containing the initial state.  The 
Hamiltonian, expressed in the canonical basis $\vert n,m\ra$,  is the sum of  
a diagonal matrix and a non-diagonal matrix.  The diagonal matrix is the sum 
of the free Hamiltonian $\hat{H}_0$ and the nonlinear term $\hat{H}_n$.  The 
corresponding matrix elements  are
\beqn
(\hat{H}_0+\hat{H}_n)_{j,k}=\left[\omega N-{\gamma\over 2}(N^2+N-2Nj+2j^2)
\right]\delta_{j,k}~~~~j,k=0,1,2,\cdott N.
\eeqn
  The interaction $\hat{H}_c$ is nondiagonal whose elements are
\beqn
(\hat{H}_c)_{j,j+1}=\sqrt{(j+1)(N-j)}~~~~j=1,2,\cdott N-1,
\eeqn
and other elements vanish.
This truncated Hamiltonian matrix is sufficient to study the evolution of states 
belonging to the 
particular invariant subspace $S_N$.  This finite size matrix makes it easier 
to numerically track the evolution of the system.  The results given 
subsequently 
are based on numerical computations.  For the calculations presented here 
the largest matrix is of order five, used to study states in the invariant 
subspace $S_4$.  For an initial condition involving states 
from two or more invariant subspaces, the Hamiltonian matrix has to be 
enlarged.  It is to be noted that the evolution dictated by $\hat{H}$ is 
solvable 
analytically\cite{korol}.  If $\aao(t)$ and $\bao(t)$ represent time-evolved 
operators, they are related to the initial operators $\aao$ and $\bao$ through 
\begin{equation}
\left(
\begin{array}{c}
\aao(t)\\
\bao(t)
\end{array}
\right)=\exp(-i{3\gamma t\over 2}\hat{N})
\left[
\begin{array}{cc}
        \exp[-i(\epsilon-{\gamma \over 2}\hat{H_i})t] + H. c. & 
\exp[-i(\epsilon-{\gamma \over 2}\hat{H_i})t] - H. c. \\
         \exp[-i(\epsilon-{\gamma t\over 2}\hat{H_i})t] - H. c. & 
\exp[-i(\epsilon-{\gamma t\over 2}\hat{H_i})t] + H. c. \\
     \end{array}
\right] 
\left(
\begin{array}{c}
\aao\\
\bao
\end{array}
\right)
\end{equation}
Here $\hat{H_i}=\aco\bao+\aao\bco$.
The term $\exp(i\gamma t\hat{H_i}/2)$ does not factorize further.  As a 
consequence, the expressions for expectation values  
become lengthy and cumbersome.  This justifies the use of numerical 
calculations which are very accurate.\\\\  
As noted earlier, states of the form $\vert N-m,m\ra$ are the only product 
states in the invariant subspace $S_N$.  
  Due to the coupling, the two modes may get entangled 
during evolution even though the initial state is a product state.
 Quantification of  
entanglement between the oscillators is easily done using  density operators. 
If the initial state of the system is in one of the invariant subspaces, say, 
$S_N$, then the  density operator for the bipartite system is
\beqn
\hat{\rho}=\sum_{m=0}^N\sum_{n=0}^N c_n c^*_m\vert N-n,n\ra\la N-m,m\vert.
\eeqn
The coefficients are time-dependent and satisfy suitable  initial conditions.  
The reduced density operators for the two oscillators are
\beqn\label{dmatA}
\hat{\rho}_a=\sum_{n=0}^N\vert c_n\vert^2\vert N-n\ra\la N-n\vert,
\eeqn
and
\beqn\label{dmatB}
\hat{\rho}_b=\sum_{n=0}^N\vert c_n\vert^2\vert n\ra\la n\vert
\eeqn
The suffixes $a$ and $b$ indicate the two modes respectively.  The reduced 
density operators are diagonal in their respective number state basis. Further, 
the diagonal elements of the reduced density operator are the same except for a 
reversal of ordering:  the coefficient $\vert c_n\vert^2$ appears as the 
probability for the state $\vert n\ra$ for the $a$-mode while it appears as 
that of the state $\vert N-n\ra$ for the other mode.  This restriction is only 
because the state of the bipartite system belongs to the subspace $S_N$.  

   A measure of 
entanglement in bipartite, pure states 
 is the linear entropy $L$ defined as follows\cite{entropy}:
\beqn
L=1-Tr\left[\hat{\rho}_a\hat{\rho}_a\right].
\eeqn
  The numerical value is independent of whether $\hat{\rho}_a$ or $\hat{\rho}_b$ 
is used in the expression.  Using the reduced density matrix expression given in 
Eq. \ref{dmatA}, the linear entropy is $1-\sum_{n=0}^N\vert c_n\vert^4$.  
Another  measure of entanglement\cite{entropy} is  von Neuman entropy $S$ defined 
as 
\beqn
S=Tr\left[\hat{\rho}_a\log\hat{\rho}_a\right].
\eeqn
  Its value is zero for separable states and reaches the maximum value of 
$\log_2 M$ for a system whose Hilbert space is of  dimension $M$. The explicit 
expression for $S$ is $2\sum_{n=0}^N\vert c_n\vert^2\log_2\vert c_n\vert$.   
Nonzero values of entropy ($S$ or $L$) imply that the system is entangled. In 
Fig. 2 the dynamics of $S$ is shown as the system evolves from initial 
conditions of the form $\vert n,m\ra$ which are separable states and hence the 
entropy is zero. Instead of plotting $S$, the ratio of $S$ and $\log_2(N+1)$ 
is plotted. The ratio varies from zero to a maximum of unity for any state.  
This makes it easier to compare the evolution of $S$ for initial states from 
different invariant subspaces.  The evolution of entropy when the total 
quantum number is one, two and three are shown in Figs. 2a -2c respectively.  
The entropy increases 
substantially approaching  the maximum attainable entanglement and subsequently 
oscillates.  However, the state $\vert 0,4\ra$, whose entropy evolution  is 
shown in Fig. 2d, does not evolve to states of high entropy.  The entropy of 
the local mode $\vert 0,4\ra$ is zero as it is a product state. Being an 
approximate eigenstate, the evolved states do not differ significantly from 
the initial state. As a consequence, the entropy remains low during evolution.  
In practical terms, this implies that local modes are not good candidates for 
generating states of high entanglement using bilinear coupling.   The situation 
changes drastically if other initial conditions are chosen.  When the system 
evolves from initial conditions other than the local modes, the entropy $S$ 
attains  values closer to the maximum allowed value of $\log_2N$.  
Evolution of entanglement when the initial states are not local modes is shown 
in Fig. 3. 
In particular, 
comparison of Fig. 2d for the state$\vert 0,4\ra$  and Fig. 3(d) for the state 
$\vert 1,3\ra$ of same total quantum number, shows the clear distinction 
between the evolution of a local and a nonlocal mode for $N = 4$.  The evolution 
takes the system to maximally entangled states whenever the reduced density operator 
of the form   ${1\over\sqrt{N+1}}\sum_{n=0}^{N}\vert n\ra\la n\vert$ during evolution.\\

     It is important to note that if the nonlinear term is absent, the  
bilinear coupling cannot lead to entangled states from separable, initial  
states unless the  initial states are nonclassical\cite{kim, wang}.  Here 
"nonclassical" implies that the Glauber-Sudarshan $P$-function for the state 
in the diagonal coherent state representation does not qualify as a probability 
density\cite{mandelwolf, gerryknight}.  The states in $S_N$ for $N>0$ are 
non-classical and hence the evolved states are entangled.   The presence of the 
nonlinearity along with  the bilinear coupling can entangle states which are 
classical.  In essence, bilinear cross coupling, without the presence of 
nonlinear term in the Hamiltonian, cannot generate entanglement if the initial 
states are classical.

\begin{figure}[htp]
\centering
\includegraphics[width=10cm, height=8cm]{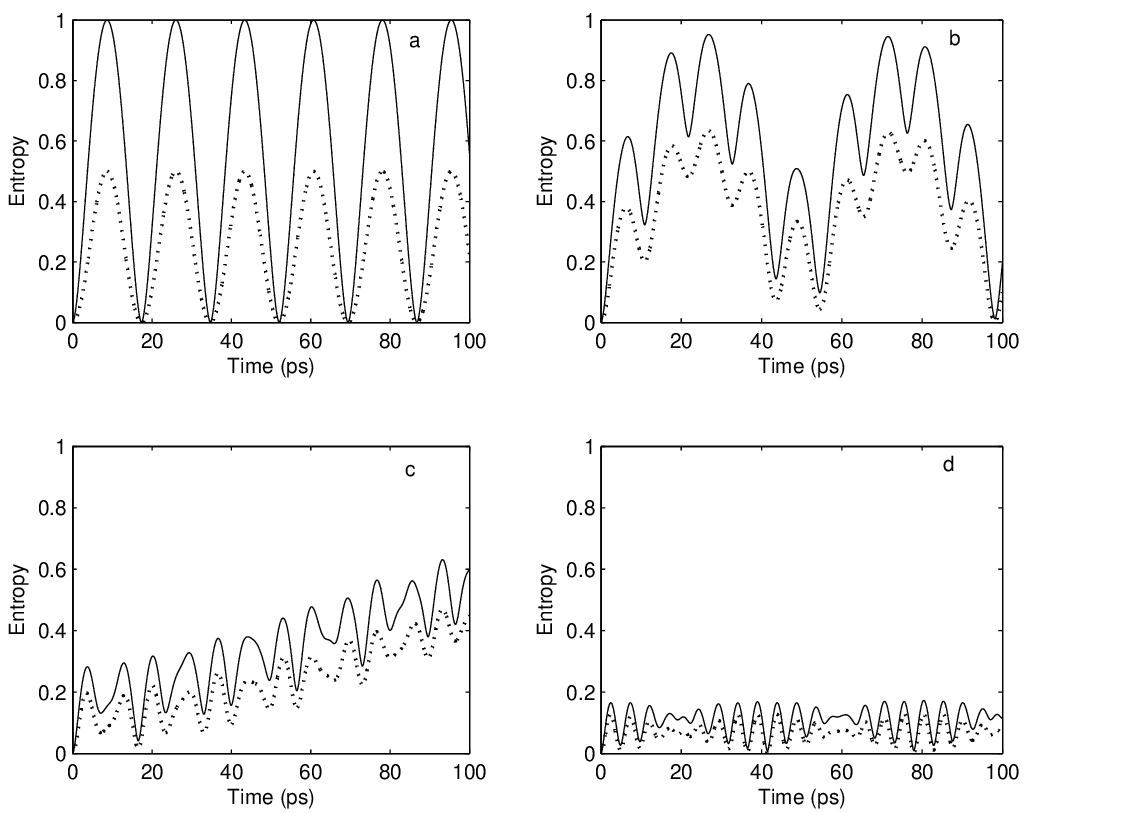}
\label{fig:Fig2}
\caption{Temporal evolution of von Neuman entropy $S$ and linear entropy $L$.  The 
quantities  plotted are $Tr[\rho_1\log\rho_1]\over\log_2(N+1)$ (continuous 
curve) 
and $L$ (dashed curve) for various initial conditions.  Figures (a)-(d) show the 
evolution as the bipartite system evolves from the initial conditions 
$\vert 0,1\ra, \vert 0,2\ra, \vert 0,3\ra$ and $\vert 0,4\ra$ respectively. 
Values of $\gamma,\epsilon$ and $\omega$ are as indicated in Fig. 1. }
\end{figure}
 \begin{figure}[htp]
\centering
\includegraphics[width=10cm, height=8cm]{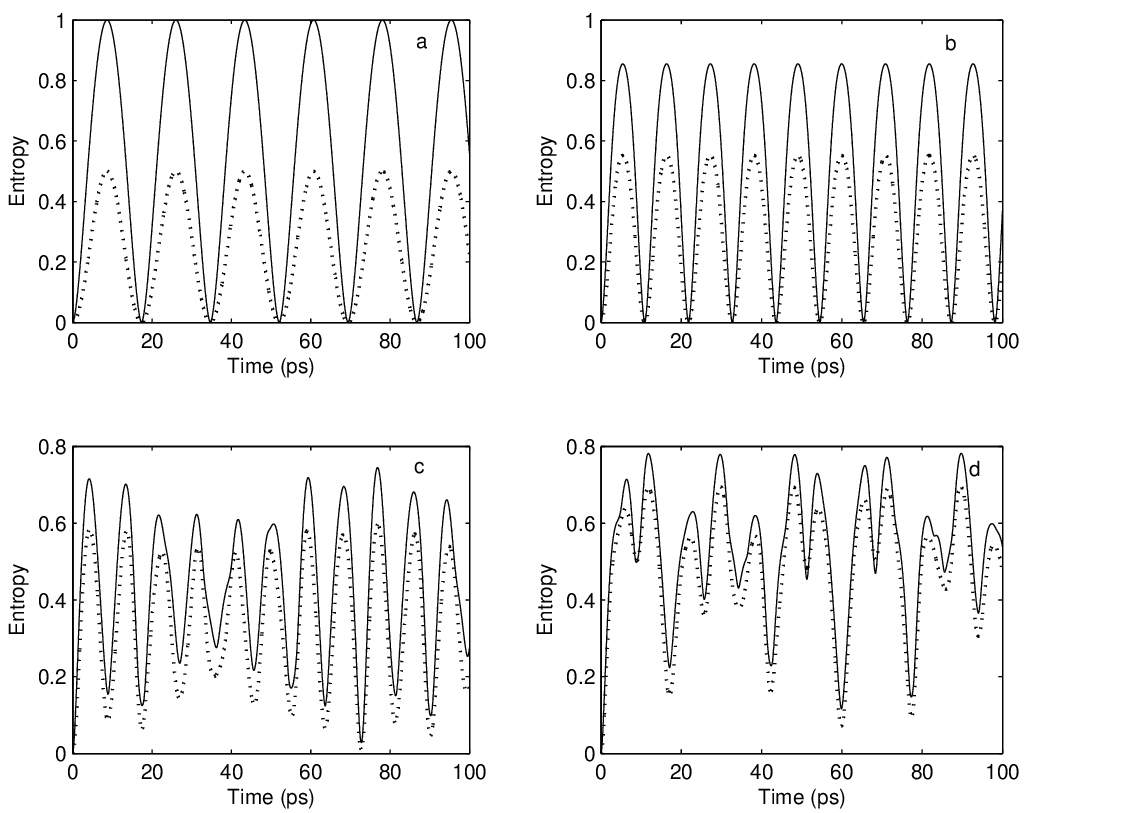}
\caption{Temporal evolution of von Neumann entropy $S$ and linear entropy $L$.  
The quantities  
plotted are $Tr[\rho_1\log\rho_1]\over\log_2(N+1)$ (continuous curve) and $L$ 
(dashed curve) for various initial conditions.  Figures (a)-(d) show the 
evolution as the bipartite system evolves from the initial conditions 
$\vert 1,0\ra, \vert 1,1\ra, \vert 1,2\ra$ and $\vert 1,3\ra$ respectively.
  For the state $\vert 1,3\ra$, the ratio attains values close to 0.8, 
indicating the entanglement is nearly 3 ebits. Higher energy states can 
generate more entanglement.  Values of $\gamma,\epsilon$ and $\omega$ are 
as indicated in Fig. 1.}
\label{fig:Fig3}
\end{figure}
\begin{figure}[htp]
\centering
\includegraphics[width=10cm, height=8cm]{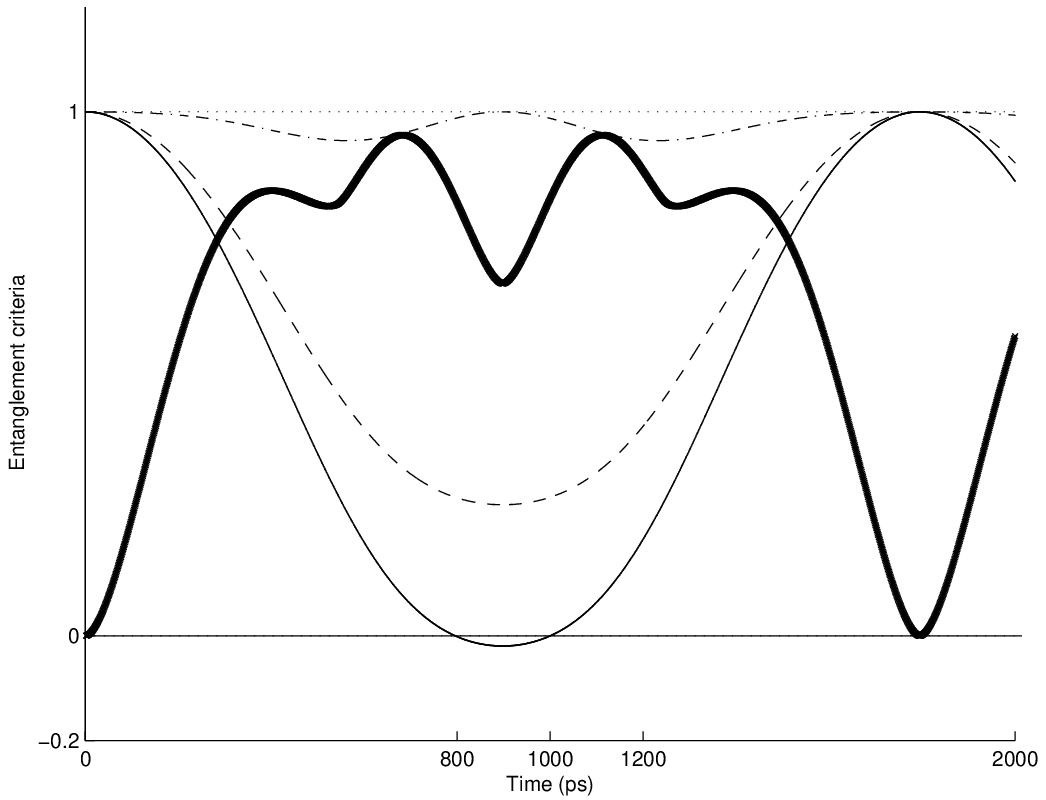}
\caption{Temporal evolution of criteria for entanglement as the system evolves 
from the initial state $\vert 2,2\ra$.  The criteria shown are 
$f_{HS}$(dashed), $f_S$ (dash-dot) and  $f_{11}$ (thin continuous). For 
comparison, the von Neumann entropy is shown (thick continuous).   The curves 
correspond to ratio of the criterion to its absolute maximum value during 
the evolution.  This quantity has been chosen so that the various criteria are 
limited to a maximum value of one.  This ratio is sufficient as it is the 
occurrence of negative value that is of importance and not actual values of the 
criteria. The value of 
$f_{11}$ falls below zero (the horizontal line) between 800 ps to 1000 ps 
indicating entanglement during those instants.  At other instants, the          
criterion 
does not detect entanglement.Values of $\gamma,\epsilon$ and $\omega$ are as    
indicated in Fig. 1. }
\label{fig:Fig4}
\end{figure}

%%%%%%%%%%%%%%%%%%%%%%%%%%%%%%%%%%%%%%%%%%%%%%%%%%%%%%%%%%%%%%%%%%%%%%%%%%%%%%%%%%%%%%%%5
The states under consideration are non-Gaussian, pure  states.   
While von Neumann entropy provides the unique measure of entanglement for pure 
states, it is required 
to have a criterion expressed in terms of measurable quantities like the 
moments of the creation-, annihilation- and number- operators. Many criteria 
have 
been proposed to detect entanglement, but none of them are universally 
applicable.  
All these criteria are sufficient conditions for entanglement.  For instance, 
the 
sufficient and necessary conditions for 
entanglement in bipartite Gaussian states are known\cite{duan, simon}.  However, 
for non-Gaussian 
states these criteria may not work.  In what follows, several of the known 
criteria are applied to the states in the invariant subspaces.   The 
explicit calculations are simplified by using the result that the expectation 
value of $\aao^m\bao^n\hat{a}^{\dagger p}\hat{b}^{\dagger q}$ in the states in 
any invariant subspace is zero if 
$m+n\ne p+q$, where $m,n,p$ and $q$ are non-negative integers. \\\\
1. Duan {\em et al} criterion \cite{duan}:  This criterion is expressed in 
terms of uncertainties in 
the two-mode operators:
\bear
\hat{u}&=&{1\over\sqrt{2}}\left[\vert m\vert (\aao+\aco)+{1\over m}(\bao+\bco)
\right],\\
\hat{v}&=&{1\over i\sqrt{2}}\left[\vert m\vert (\aao-\aco)-{1\over m}
(\bao-\bco)\right].
\eear
Separable states satisfy $\la(\Delta u)^2\ra+\la(\Delta v)^2\ra > (1+m^4)/m^2$. 
 For the states in $S_N$, the relation becomes 
$\la(\Delta u)^2\ra+\la(\Delta v)^2\ra=
(\vert m\vert^4+1)/m^2+\vert m\vert^2N_a+ N_b/m^2$, where $N_a$ and $N_b$ refer 
to the expectation values $\la\ano\ra$ and $\la\bno\ra$ respectively.  Since 
the number operators 
are non-negative operators, the sum of the uncertainties indeed satisfies the 
condition of separability although the states are entangled as evident from the  
earlier discussion on the von Neumann entropy.  Hence this criterion is not 
useful in the present context.\\\\
2. Mancini {\it et al} criterion : This criterion employs the product of 
uncertainties in the operators considered in the criterion due to 
Duan {\it et al}, choosing $m=1$\cite{mancini}.  
The inequality to be satisfied to identify entangled states is that
\beqn
\la(\Delta u)^2\ra\la(\Delta v)^2\ra  < 1.
\eeqn
When applied to the states in $S_N$, 
$\la(\Delta u)^2\ra = 1+\la\ano+\bno\ra+\la\aco\bao+\aao\bco\ra$ which can be 
written as $1+\la(\aco+\bco)(\aao+\bao)\ra$.  The operator 
$(\aco+\bco)(\aao+\bao)$ is of the form $\hat{A}^\dagger\hat{A}$, which is 
positive.  Hence, $\la(\Delta u)^2\ra >1$.  Similarly,  
$\la(\Delta v)^2\ra = 1+\la(\aco-\bco)(\aao-\bao)\ra  > 1$.  Thus, the product 
of the uncertainties is larger than unity making the criterion not useful in 
identifying the entangled states in $S_N$.\\\\  
3. Shchukin-Vogel's criteria:\\
Many of the known criteria are expressible in terms of the determinants 
introduced in \cite{vogel}:\\ 
All separable states 
satisfy 
\begin{equation}
D_3=\left|
\begin{array}{ccc}
        {1} & {\la\aco\ra} & {\la\bco\ra} \\
        {\la\aao\ra} & {\la\ano\ra} &{\la\aco\bco\ra}\\
{\la\bao\ra} & {\la\aao\bao\ra} &{\la\bco\bao\ra}
     \end{array} 
\right|\ge0.
\end{equation}
This inequality is a stronger version of Duan {\it et al} criterion.
 If the determinant 
is negative, the state is entangled.  For the states in an invariant 
subspace 
$S_N$, the determinant is explicitly evaluated to give $\la\ano\ra\la\bno\ra$ 
which is  positive.  Thus, the criterion does not detect entanglement in these 
states. \\
Another criterion that works for  entangled coherent states 
$\vert\alpha\ra
\vert\beta\ra-\vert-\alpha\ra\vert-\beta\ra$  is that  
\begin{equation}
\left|
\begin{array}{ccc}
        {1} & {\la\bao\ra} & {\la\aao\bco\ra} \\
        {\la\bco\ra} & {\la\bno\ra} &{\la\aco\bno\ra}\\
{\la\aco\bao\ra} & {\la\aco\bco\bao\ra} &{\la\ano\bno\ra}
     \end{array} 
\right| < 0,
\end{equation}
implies entanglement.  For the states under consideration in the present work, 
the determinant is  
$\la\ano\bno\ra\la\bno\ra$ which is positive and hence inconclusive about the 
entanglement in the states.\\\\
4. SU(2) criterion: The operators corresponding to the two modes can be 
combined 
to obey SU(2) and SU(1,1) algebra.  The corresponding uncertainty relations 
provide 
criteria to detect entanglement.  The expression for SU(2) uncertainty product 
satisfied by separable states \cite{agarwal,nha} when applied to the states 
in the invariant subspaces is   
\begin{equation}
\la\ano\bno\ra\left[4\la\ano\bno\ra+2\la\ano\ra+2\la\bno\ra\right]+
4\la\ano\ra\la\bno\ra > 0.  
\end{equation}
However, to detect entanglement this expression should attain negative values.  
Since the expression is  positive, it does not detect entanglement in the 
states.\\\\
5. It is known that SU(1,1) \cite{agarwal, nha} uncertainty relation can detect 
entanglement in a 
class of two-mode non-Gaussian states involving the ground and the first 
excited states of harmonic oscillator. For the states in any of the invariant  
subspaces, the criterion states that if 
\bear
f_{11} &=& \left[1+2N_{ab}+N_a+N_b-2N_aN_b\right]^2-4\left[\Re\la\aaot\bcot\ra^2
-\Re\la\aco\bao\ra^2\right]^2\nonumber\\
& &-\vert N_a+N_b\ra\vert^2 < 0,
\eear  
the corresponding state is entangled.  Here $N_{ab}$ is  the 
expectation value $\la\ano\bno\ra$ and $\Re$ stands for real part of the 
expression that follows.\\   
Simiarly, Simon's criterion\cite{simon} identifies entangled states in the 
invariant subspaces if 
\bear
f_S&=&{\Re\la\aao\bco\ra^2\over 2}+\vert\la\aao\bco\ra\vert^4-
\vert\la\aao\bco\ra\vert^2\left[N_a+N_b+2N_aN_b\right]\nonumber\\
& &+{1\over16}(1+2N_a)^2(1+2N_b)^2 < 0. 
\eear  
The criterion due to Hillery and Zubairy \cite{hillery} states that if  
\beqn
f_{HZ}= N_aN_b-\vert\la\aco\bao\ra\vert^2 < 0, 
\eeqn
then the state is entangled.  
For the states generated when the system evolves from the initial condition 
$\vert 2,2\ra$, these three criteria  are evaluated numerically.  The results 
are shown in 
Fig.\ref{fig:Fig4}.    It is seen that the entanglement is not identified by 
the criteria $f_{HZ}$ and $f_S$ .  The SU(1,1) criterion 
detects entanglement in at least some of the states occurring during evolution.  
This is the 
reason for choosing the $\vert2,2\ra$ as the initial state.  If any other 
product state in $S_4$  is chosen as the initial condition, all the three 
criteria fail.   

\begin{figure}[htp]
\centering
\includegraphics[width=10cm, height=8cm]{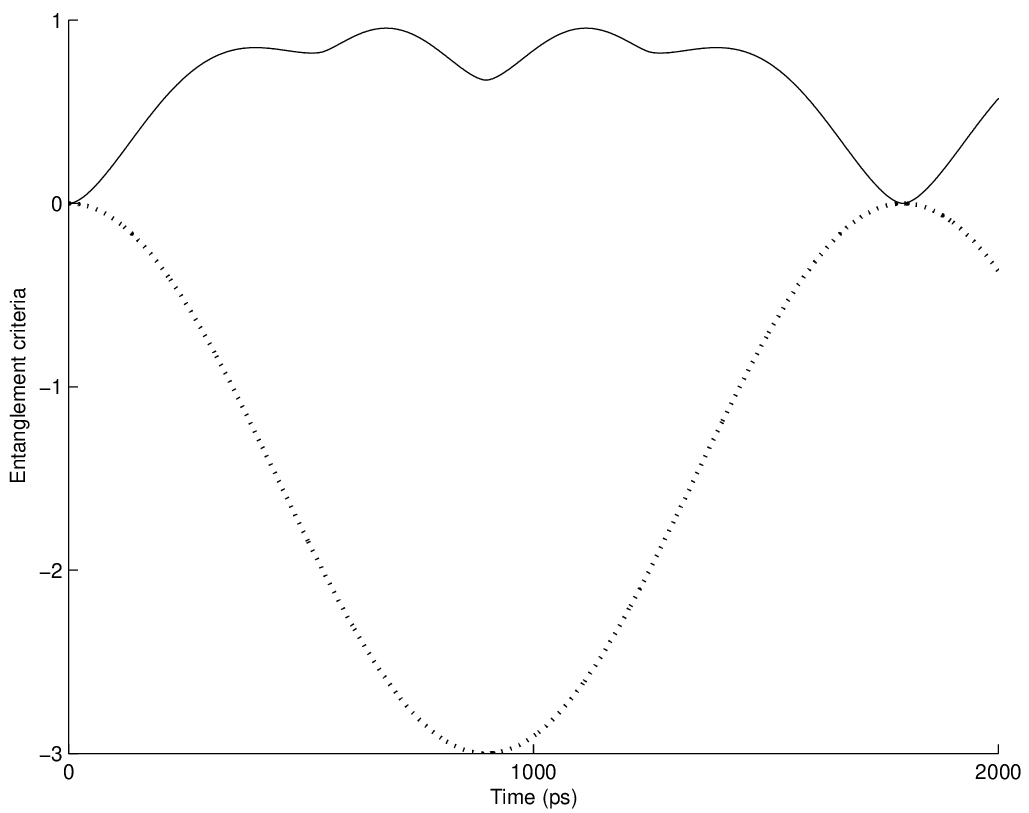}
\caption{Temporal evolution of criteria for entanglement as the system evolves   
from the initial state $\vert 2,2\ra$.   The quantities plotted are 
von Neumann entropy (conitnuous) and $D$ (dashed).   
Values of $\gamma,\epsilon$ and $\omega$ are as indicated in Fig. 1. }
\label{fig:Fig5}
\end{figure}

 The failure of the known criteria to detect entanglement provides motivation 
to look for a new criterion.   The states in an invariant subspace are spanned 
by states of fixed 
total quantum number $N=n+m$, leading to  strong correlation between the 
number of quanta of the first mode and that of the second mode.  It is, 
therefore, natural to expect that a correlation function of the number operators 
of the two modes may detect entanglement.  Such a  criterion for 
entanglement is derived as follows.  For all product 
states the equality $\la\ano\bno\ra=\la\ano\ra\la\bno\ra$ holds.  Consider 
the most general state $\sum c_n\vert n,N-n\ra$, where the summing index $n$ 
runs from zero to $N$.  For these states, 
$\la\ano\bno\ra=N\la\ano\ra-\la\ano\ano\ra$ and 
$\la\ano\ra\la\bno\ra=N\la\ano\ra-\la\ano\ra^2$.  But, 
$\la\ano\ano\ra\ge\la\ano\ra^2$ 
for any state.  Comparing the expressions for $\la\ano\bno\ra$ and $
\la\ano\ra\la\bno\ra$ yields the inequality 
$\la\ano\bno\ra<\la\ano\ra\la\bno\ra$.  The equality sign holds if 
$\la\ano\ano\ra=\la\ano\ra^2$, which is true if all but 
one of the $c_n$s is zero.  In that case the state is a product state and 
there is no entanglement.  To bring out these features, the temporal 
variation of the number correlation function 
 $D=\la\ano\bno\ra-\la\ano\ra\la\bno\ra$ is shown 
in Fig.\ref{fig:Fig5}.  Negative value of $D$ implies that the state is 
entangled.  
For comparison, the von Neumann entropy is shown.  It is seen that whenever 
the entropy is positive, which indicates entanglement, the quantity $D$ is 
negative.  Thus, $D$ qualifies as a suitable criterion to detect entanglement 
in the class of non-Gaussian, pure states considered.   

\section{Quadrature variances}\label{secV}
The coupled C-H oscillators modelled by $\hat{H}$ are oscillating masses.  
The quadratures of these oscillators are naturally identified with their 
positions  and momenta.  For the individual oscillators described by $\hat{H}$,  
the position and momentum quadratures  are given by 
\bear
\hat{Q}_a&=&{\aao+\aco\over\sqrt{2}},\\
\hat{P}_a&=&{\aao-\aco\over i\sqrt{2}},
\eear
respectively and they satisfy $[\hat{Q}_a,\hat{P}_a]=i$, where $\hbar$ is set 
equal to unity. The corresponding variances satisfy 
$2\Delta Q_a\Delta P_a\ge 1$.  
If the variance of one of the quadratures is less than 1/2, the corresponding 
quadrature is said to exhibit squeezing.  Evaluation of the variances in the 
present case is simplified since the expectation values of operators involving 
creation and annihilation operators with unequal exponents,  vanish in the 
states within an invariant subspace.  For instance, the expectation value of 
$\aao\acot$ is zero as creation and annihilation operators occur with different 
exponents;  similarly  the expectation values of $\aao\bao, \aco\bco$, etc all 
vanish for the states in $S_N$. Hence, for the states belonging to $S_N$ ,
\beqn
\Delta Q_a =\sqrt{{1\over 2}[1+N]},
\eeqn
and 
\beqn
\Delta P_a=\sqrt{{1\over 2}[1+N]}.
\eeqn
The suffix $a$ labels the mode.  Similar expressions can be written for the 
other mode and the results on variances are identical to those of the $a-$mode.  
If  $N > 0$ the quadratures of the individual oscillators do not exhibit any 
squeezing for the states in the invariant subspaces. Further, the numerical 
values of the variances in the two quadratures are equal and increases with the
 total quantum number $N$.  The state $\vert 0,0\ra$ does not exhibit squeezing 
and  it corresponds to minimum uncertainty state.  The notion of quadratures 
associated with the individual modes has been generalized to multi-mode cases.   
For the bipartite system, the two-mode quadratures are defined as\cite{caves} 
\bear
\hat{d}_1&=&{\aao+\bao+\aco+\bco\over 2^{3\over2}},\\
\hat{d}_2&=&{\aao+\bao-\aco-\bco\over i2^{3\over2}}.\\
\eear
Based on these expressions for the two-mode quadratures, the respective 
variances 
of the quadratures in any state belonging to $S_N$ are given by
\bear
(\Delta d_1)^2&=&{1\over 4}\left[1+N+\la\aao\bco\ra+\la\aco\bao\ra\right],\\
(\Delta d_2)^2&=&{1\over 4}\left[1+N+\la\aao\bco\ra+\la\aco\bao\ra\right].
\eear
The states  belong to the invariant 
subspace $S_N$ possess equal two-mode quadrature variances. The two-mode 
quadratures satisfy the commutation relation $\left[d_1,d_2\right]=i/2$ and 
the corresponding uncertainty relation is $\Delta d_1\Delta d_2\ge1/4$.  If 
any of the quadrature has uncertainty  lower than $1\over 2$, the quadrature 
is said to be squeezed.  The fact that the two-mode variances are equal 
implies that one of the variances cannot be reduced to values less than $1/2$, 
the condition for squeezing, without violating the uncertainty relation.  In 
short, squeezed fluctuations in the single-mode or two-mode  quadratures  are
 not possible for the states in any of the invariant subspaces. Though there is 
no squeezing, the states in the invariant subspaces  are very much nonclassical 
in the sense that their respective $P$-distributions are highly singular.\\

\section{Generation of maximally entangled states}\label{secVI}

	Bell states enjoy a special status in quantum information theory as they 
are maximally entangled bipartite states\cite{nielsen}.  The canonical Bell 
states are the linear superpositions $\vert 0,0\ra\pm\vert 1,1$ and 
$\vert 0,1\ra\pm\vert 1,0\ra$.  Bell-like states are similar combinations 
allowing for a relative phase between the superposed states.  Though the C-H 
oscillators are not two level systems, the presence of invariant subspaces 
allows for the creation of Bell-like states for the two coupled oscillators.  
This special feature is available in the invariant subspace $S_1$ spanned by 
$\vert 0,1\ra$ and $\vert 1,0\ra$. 
    The Hamiltonian $\hat{H}_2$ required to 
describe the dynamics of the states in the subspace $S_2$ is
%\beqn
$$\hat{H}_2 =
\left[
\begin{array}{cc}
        {\omega-\gamma} & {-\epsilon} \\
        {-\epsilon} & {\omega-\gamma}
     \end{array} 
\right].$$
%\eeqn
The eigenvalues of the $\hat{H}_2$ are $\omega-\gamma\pm\epsilon$ and the 
corresponding eigenstates are the Bell states 
$\vert \pm\ra=\left[\vert0,1\ra\pm\vert 1,0\ra\right]/\sqrt{2}$. 

\begin{figure}[htp]
\centering
\includegraphics[width=10cm, height=8cm]{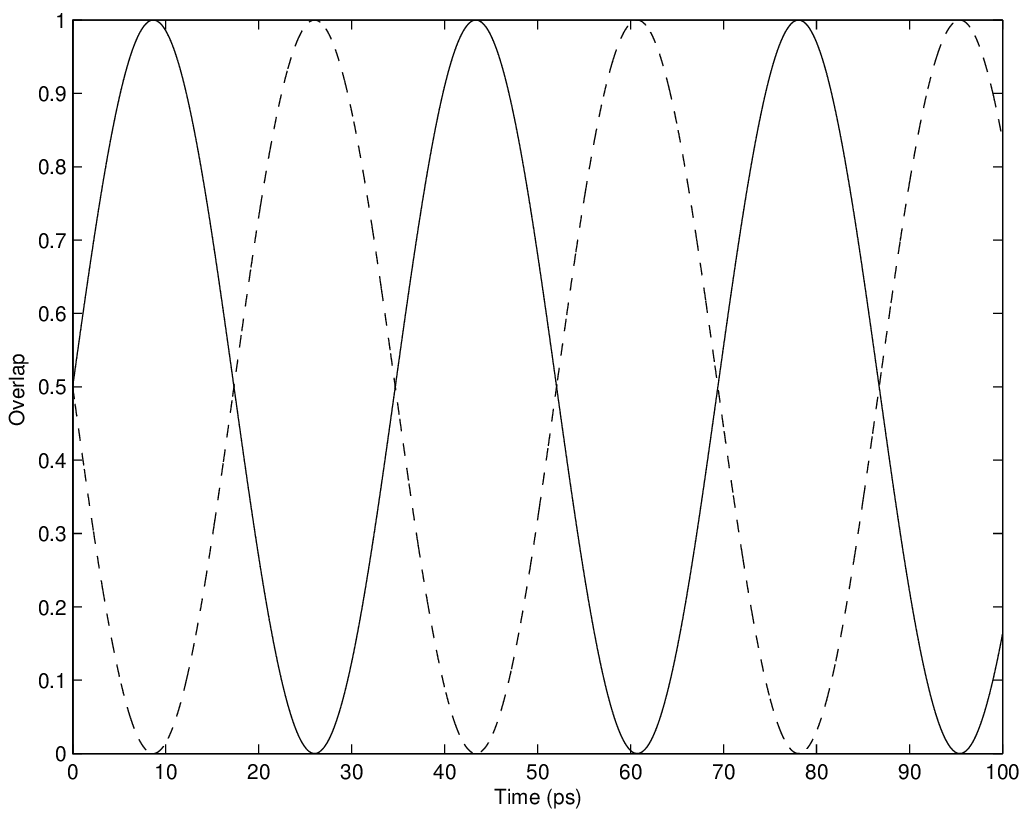}
\caption{Overlap between the state $\exp(-it\hat{H}_2)\vert 0,1\ra$ and the 
Bell-like states $\vert 0,1\ra+i\vert 1,0\ra$ (continuous) and 
$\vert 0,1\ra-i\vert 1,0\ra$ (dashed).  The instants when the overlap becomes 
unity, the initial state has evolved into the corresponding Bell-like state. 
Values of $\gamma,\epsilon$ and $\omega$ are the same as for Fig. 1}
\label{fig:Fig6}
\end{figure}

If the initial state is $\vert 0,1\ra=[\vert +\ra +\vert-\ra]/\sqrt{2}$ then 
\beqn
\exp(-it\hat{H}_2)\vert 0,1\ra \propto(1+\exp(i\epsilon t))\vert 0,1\ra+
(1-\exp(i\epsilon t))\vert 1,0\ra,
\eeqn 
apart from an overall multiplicative normalization factor.  If the time $t$ is 
chosen to be $\epsilon t=\pi/2 $, then the time-evolved state is the Bell-like 
state $\vert 0,1\ra-i\vert 1,0\ra$, apart from an overall phase factor.  If 
$\epsilon t=3\pi/2$, the initial state evolves to become another Bell-like state 
$\vert 0,1\ra+i\vert 1,0\ra$, but for an overall phase factor. The overlap 
between the time-evolved  state of the system and the Bell-like states are 
given in Fig. \ref{fig:Fig6}.  At specific instants, the initial state evolves 
to have unit 
overlap with $\vert 0,1\ra+i\vert 1,0\ra$; there is an overall phase factor to 
the actual state achieved and the overlap is insensitive to such factors.  Since 
the subspace $S_1$ is invariant under the unitary evolution, the other two 
Bell-like states, $\vert 0,0\ra\pm\vert 1,1\ra$ are not attainable with any  
initial condition contained in $S_1$.  These states belong to the direct sum of 
$S_0$ and $S_2$. Without external control fields it is not feasible to generate 
all the Bell-like states using the inherent coupling in the system.

\section{Summary}\label{results}

	The Hamiltonian for the coupled, nonlinear C-H oscillators of 
dihalomethanes  is similar to that of the nonlinear couplers with negative 
$\chi^2$. Thus the C-H oscillators are the microscopic nonlinear couplers for 
bosonic fields of vibrational-cum-bending  motion.  Like the 
Kerr nonlinear 
couplers, the dihalomethane can serve as a system wherein entanglement is 
easily generated with the advantage that there is no need to have external 
fields for the purpose of entangling the modes.   The coupling due to the 
bending motion of the molecule creates  entanglement between the modes and 
allows for exchange of energy.  The structure of the Hamiltonian with commuting 
terms allows to split the Hilbert space into irreducible, invariant subspaces.  
Each of the invariant subspace is the span of the product states of the form 
$\vert n,m\ra$ with $n+m$ fixed for the individual subspaces.  Apart from the
canonical basis states, the remaining states in the invariant subspaces are 
entangled.  The states belonging to the invariant subspaces are 
non-Gaussian states.  The know detection criteria for entangled states do 
not identify entanglement in the states belonging to the invariant subspaces. 
However, for these  non-Gaussian pure states  
the entanglement criterion is that $\la\ano\bno\ra<\la\ano\ra\la\bno\ra$.   

 The local modes of sufficient energy  are approximate eigenstates of 
the complete Hamiltonian.  The entanglement in the evolved state is much 
larger if the initial state is 
different from a local mode.  Though there is a change of quadrature variance 
during evolution, the states in the invariant subspaces do not exhibit 
single-mode or two-mode quadrature squeezing.  The variances are always above 
the minimum allowed limit. The two eigenstates of the total Hamiltonian, 
belonging to   the subspace $S_1$ correspond to two of the Bell states.  Though 
states of high entanglement cannot be generated from the local modes of high 
energy, the Hamiltonian generates maximally entangled states  during 
the evolution of the local modes $\vert 0,1\ra$ or $\vert 1,0\ra$.

%%%%%%%%%%%%%%%%%%%%    %%%%%%%%%%%%%%%%%%%%%%%%%%%%%%%%%%%%%%%%%%%%%%%%%%%%%%%%


\begin{thebibliography}{40}
\bibitem{nielsen}{M. A. Nielsen and I. L. Chuang  {\it Quantum Computation and 
Quantum Information} (Cambridge: Cambridge University Press) (2000)}
\bibitem{devices}{G. Chen {\it et al} {\it Quantum Devices Principles, 
Design, and Analysis} (USA: Chapman \& Hall)(2007)}
\bibitem{tesch}{C. M. Tesch, L. Kurtz and R. de Vivie-Riedle {\it Chemical 
Physics Letters} {\bf 343}(2001) 633}
\bibitem{babikov}{D. Babikov {\it Journal of Chemical Physics} {\bf 121} 
(2004) 7577}
\bibitem{scott}{A. S. Scott {\it Nonlinear Science: Emergence and 
Dynamics of Coherent Structures}, (London: Oxford University Press)(2003)}
\bibitem{chefles}{A. Chefles  and S. M. Barnett {\it Journal of Modern Optics} 
{\bf 43} (1996) 709}
\bibitem{bernstein}{L. J. Bernstein {\it Physica} D {\bf 68} (1993) 174}
\bibitem{bernsteinRMP}{S L Braunstein and P van Loock {\it Reviews of 
Modern Physics} {\bf 77} (2005) 513}
\bibitem{furusawa}{A. Furusawa {\it et al} {\it Science} {\bf 282} (1998) 706}
\bibitem{zanardi}{E. Ciancio and P. Zanardi {\it Physics Letters A} 
{\bf 360} (2006) 49}
\bibitem{perina}{J. Perina {\it Quantum Statistics of linear and nonlinear 
optical phenomena} (The Netherlands: Kluwer Academic Publishers)(1991)}
\bibitem{agarwalpuri}{G. S. Agarwal and R. R. Puri {\it Physical Review A} 
{\bf 39} (1989) 2969}
\bibitem{jensen}{S. M. Jensen {\it IEEE J. Quantum Electron.} {\bf QE-18} 
(1982) 1580}  
\bibitem{miranowicz_jpb}{A. Miranowicz and W. Leonski {\it Journal of 
Physics B: Atomic, Molecular and Optical Physics} {\bf 39} (2006) 1683}
\bibitem{leonski_jopb}{W. Leonski and A. Miranowicz {\it Journal of Optics B}
 {\bf 6} (2004) S37.}
\bibitem{sanz_jpa}{L. Sanz, R. M. Angelo and K. Furuya {\it Journal of Physics 
A: Mathematical and General} {\bf 36} (2003) 9737}
\bibitem{ariunbold}{G. Ariunbold and J. Perina {\it Journal of Modern Optics} 
{\bf 48} (2001) 1005}
\bibitem{olsen}{M. K. Olsen {\it Physical Review A} {\bf 73} (2006) 053806}
\bibitem{sudeesh}{C. Sudheesh, S. Lakshmibala and V. Balakrishnan {\it Journal 
of Physics B: Atomic,Molecular and Optical Physics} {\bf 39} (2006) 3345}
\bibitem{kowalewska}{A. Kowalewska-Kudlaszyk and W. Leonski {\it Physical 
Review A} {\bf 73} (2006) 042318}
\bibitem{entropy}{C. H. Bennett, H. J. Bernstein, S. Popescu and B. Schumacher 
{\it Physical Review A} {\bf 53} (1996) 2046}
\bibitem{kim}{M. S. Kim, W. Son, V. Buzek and P. L. Knight {\it Physical Review 
A} {\bf 65} (2002) 032323}
\bibitem{wang}{Wang Xiang-bin {\it Physical Review A} {\bf 66} (2002) 024303}
\bibitem{mandelwolf}{L. Mandel and E. Wolf {\it Optical Coherence and 
Quantum Optics} (Cambridge: Cambridge University Press)(1998)}
\bibitem{korol}{N. Korolkova and J. Perina {\it Optics Communications} {\bf 136} 
(1996) 135}
\bibitem{gerryknight}{G. C. Gerry and P. L. Knight {\it Introductory Quantum 
Optics} (Cambridge: Cambridge University Press) (2004)}
\bibitem{duan}{L.-M. Duan {\em et al} {\it Physical Review Letters} {\bf 84} 
(2000) 2722}
\bibitem{simon}{R. Simon {\it Physical Review Letters} {\bf 84} (2000) 2726}
\bibitem{mancini}{S. Mancini {\em et al} {\it Physical Review Letters} {\bf 88} 
(2002) 120401}
\bibitem{vogel}{E. Shchukin and W. Vogel {\it Physical Review Letters} 
{\bf 95} (2005) 230502}
\bibitem{agarwal}{G. S. Agarwal and A. Biswas {\it New Journal of Physics} 
{\bf 7} (2005) 211}
\bibitem{nha}{H. Nha and J. Kim {\it Physical Review A} {\bf 74} (2006) 012317}
\bibitem{hillery}{M. Hillery and M. S. Zubairy {\it Physical Review Letters} 
{\bf 96} (2006) 050503}
\bibitem{caves}{C. M. Caves {\it Physical Review D} {\bf 26} (1985) 1817}

\end{thebibliography}
\end{document}